\documentclass[conference]{IEEEtran}
\IEEEoverridecommandlockouts
\usepackage{cite}
\usepackage{amsmath,amssymb,amsfonts}
\usepackage{graphicx}
\usepackage{textcomp}
\usepackage{xcolor}
\usepackage{algorithmicx,algorithm}
\usepackage{algpseudocode}
\newtheorem{definition}{Definition}
\usepackage{algpseudocode}
\usepackage[english]{babel}
\usepackage{subcaption}
\newcommand{\name}{CARES}
\newcommand{\nameinitem}{CARES }
\usepackage{url}
\def\BibTeX{{\rm B\kern-.05em{\sc i\kern-.025em b}\kern-.08em
    T\kern-.1667em\lower.7ex\hbox{E}\kern-.125emX}}
\begin{document}

\title{Context-aware Session-based Recommendation with Graph Neural Networks\\
\thanks{This work is supported by Shanghai Pujiang Talent Program No. 21PJ1402900, Shanghai Science and Technology Committee General Program No. 22ZR1419900 and National Natural Science Foundation of China No. 62202172.}
}

\author{
\IEEEauthorblockN{1\textsuperscript{st} Zhihui Zhang}
\IEEEauthorblockA{\textit{School of Data Science and Engineering} \\
\textit{East China Normal University}\\
Shanghai, China \\
51215903035@stu.ecnu.edu.cn}
\and
\IEEEauthorblockN{2\textsuperscript{rd} Jianxiang Yu}
\IEEEauthorblockA{\textit{School of Data Science and Engineering} \\
\textit{East China Normal University}\\
Shanghai, China \\
jianxiangyu@stu.ecnu.edu.cn}
\and
\IEEEauthorblockN{3\textsuperscript{nd} Xiang Li\IEEEauthorrefmark{1}\thanks{*Corresponding author}
}

\IEEEauthorblockA{\textit{School of Data Science and Engineering} \\
\textit{East China Normal University}\\
Shanghai, China \\
xiangli@dase.ecnu.edu.cn}
}

\maketitle

\begin{abstract}
Session-based recommendation (SBR) is a task that aims to predict items based on anonymous sequences of user behaviors in a session.
While there are methods that leverage rich context information in sessions for SBR, most of them
have the following limitations:
1) they fail to distinguish the item-item edge types when constructing the global graph for exploiting cross-session contexts;
2) they learn a fixed embedding vector for each item, 
which lacks the flexibility to reflect the variation of user interests across sessions;
3) they generally use the one-hot encoded vector of the target item as the hard label to predict, thus failing to capture the true user preference.
To solve these issues, we propose \name, a novel context-aware session-based recommendation model with graph neural networks,
which utilizes different types of contexts in sessions to capture user interests. 
Specifically, we first construct a multi-relation cross-session graph to connect items according to intra- and cross-session item-level contexts.
Further, to encode the variation of user interests, we design personalized item representations.
Finally, we employ a label collaboration strategy for generating soft user preference distribution as labels.
Experiments on three benchmark datasets demonstrate that \nameinitem consistently outperforms state-of-the-art models in terms of P@20 and MRR@20.
Our data and codes are publicly available at \url{https://github.com/brilliantZhang/CARES}.
\end{abstract}


\section{Introduction}
Recommendation systems play a crucial role in various fields because they provide users with personalized information to complete a task in the midst of a large amount of information.
At present, many recommendation models have achieved great success, but most of them usually need to use user profiles. 
However, as the number of users on the platform grows and privacy awareness increases, 
user profiling may not be available in certain applications.
Without obtaining user profiles as well as long-term historical user behaviors, 
it is hard to accurately model portraits of users. 
Consequently, 
session-based recommendation (SBR) has recently attracted more attention. 
Here,
a session 
can generate 
interactive behavior sequence (e.g., clicks in e-commerce scenarios) 
in a short period of time,
and SBR aims to 
predict the \emph{next item} based on an anonymous short-term behavior sequence.

\begin{figure}[t]
  \centering
\includegraphics[scale=0.25]{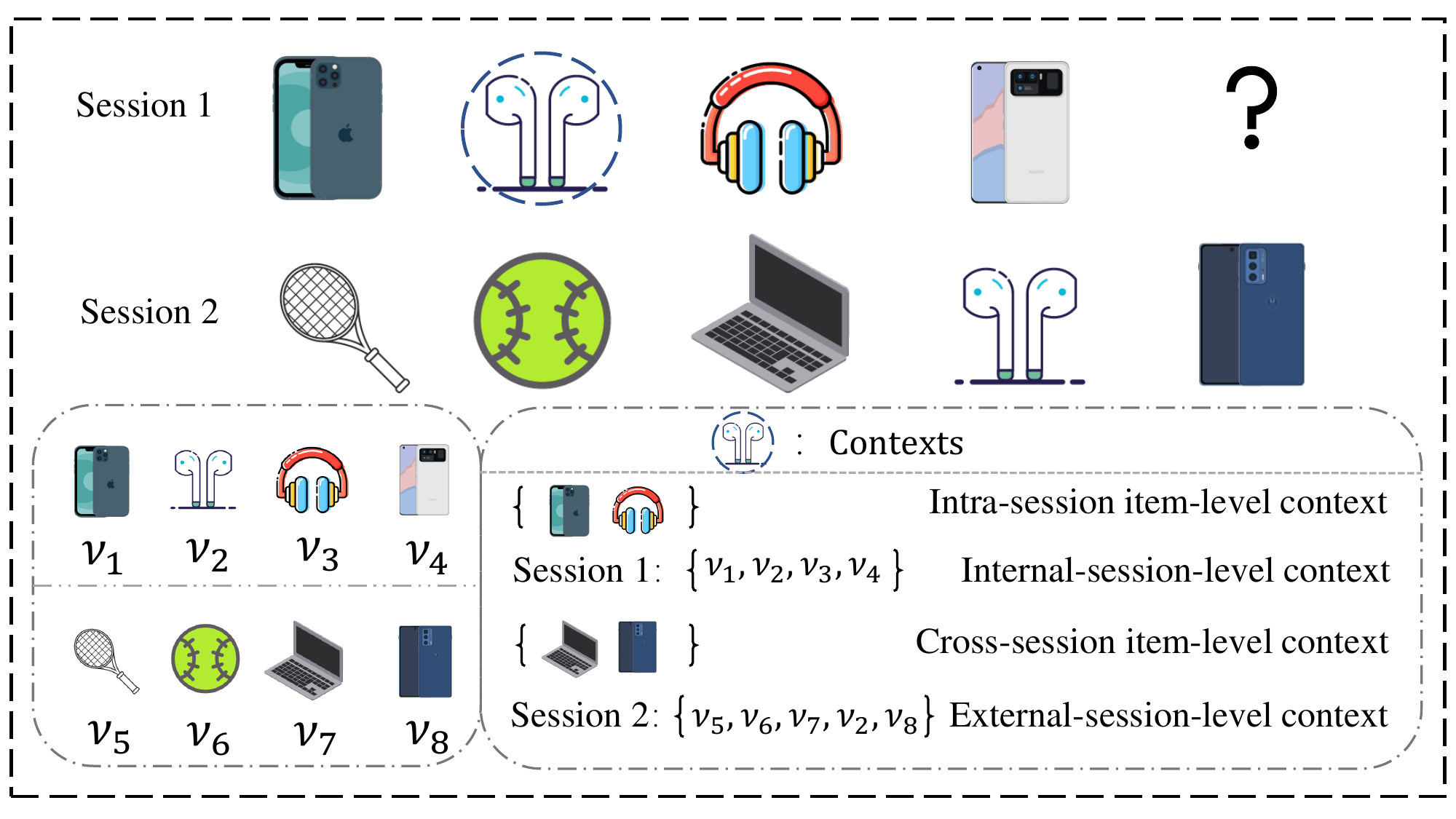}
  \caption{Contexts referred to in this paper.}
  \label{fig:intro}
\end{figure}

To address the SBR problem,
some existing methods \cite{Wang19,Wang20,Luo20,Ye20}
utilize the rich context information in sessions,
which generally includes both \emph{intra-session} and 
\emph{cross-session} ones.
For the former,
we can further divide it into 
\emph{item-level context},
which characterizes the neighboring items in the behavior sequence for an item,
and \emph{session-level context},
which refers to the complete sequence information in a session.
Similarly,
the latter includes collaborative information from sessions with similar behavioral patterns for both item and session.
Details on the division of contexts are given in Figure~\ref{fig:intro}.
Early studies \cite{Rendle10,Hidasi16,Li17} for SBR
employ intra-session contexts only,
whose performance could be adversely affected when the behavior sequence in a session is very sparse.
Recently, 
there are also methods \cite{Wang19,Wang20,Luo20,Ye20} that leverage both intra-session and cross-session contexts, 
which aim to incorporate contextual information from relevant sessions to enrich the representation of a given session. 
In particular,
some methods \cite{Wang20,Ye20} propose to construct a global graph to link items from various sessions according to the
intra- and cross-session item-level context information,
and then learn item/session embeddings based on the graph.
Despite the success,
most of these methods suffer from three major limitations.
First,
when constructing the global graph,
they fail to distinguish the item-item edge types.
Since
it has been verified in~\cite{Quadrana16,Zhou20} that integrating item attributes can improve the recommendation performance,
the categorical attributes of items (e.g., ``shirts'' and ``pants'' belong to the apparel category), 
can be used to distinguish item relations.
For example, 
if products in two categories are frequently interacted by users,
the relation between the two item categories is of more importance.
Second,
they learn a fixed embedding vector for each item.
However,
since user interest could vary across sessions,
the embedding of an item 
should be learned to reflect the variation of user interests and personalized w.r.t. different sessions.
Third,
they generally 
use the one-hot encoded vector of the target item as the hard label to be predicted, which may not reflect the true user preference.
However, the true distribution of user preferences is usually unknown, as only a limited number of items are exposed to users.
Simply regarding the one-hot encoded vector of the target item as the true distribution 
could induce bias and lead to the overfitting problem \cite{Pan22}. 

In this paper,
to address these problems, 
we propose a novel context-aware session-based recommendation model \name,
which leverages the four types of contexts introduced earlier. 
Specifically,
we first construct a multi-relation cross-session graph 
to connect items according to intra- and cross-session item-level contexts,
where edge relations are defined based on item categories.
Then
based on the graph,
we learn general item embeddings with graph neural networks (GNNs).
Further,
to encode the variation of user interests,
we also learn personalized item representations w.r.t. sessions with a gating mechanism. 
After that,
we unify item embeddings with item positions and session length to learn session representations.
Finally,
to alleviate the bias induced by the hard label of one-hot encoded vector of the target item in a session,
we employ internal- and external-session-level contexts,
and 
present a label collaboration strategy,
which uses most similar historical sessions to the current session
for collaborative filtering
and generates soft label of user preferences to be predicted.
We next summarize our main contributions in this paper as follows:
\begin{itemize}
\item We propose a novel context-aware session-based recommendation model \name.

\item We design personalized item embeddings w.r.t. sessions to capture the variation of user interests across sessions. 

\item We propose a simple and effective label collaboration method that generates soft user preference distribution as labels.

\item We conduct extensive experiments on three public benchmark datasets to show the superiority of our method over other state-of-art models.
\end{itemize}

\section{RELATED WORK}
\label{sec:related}

\subsection{Session-based Recommendation}
Early studies \cite{Sarwar01} on SBR 
use the similarity between the last item of the session and candidate items to make recommendations.
However, they omit
the sequential information in the session.
While the Markov-chain-based method \cite{Rendle10} can bridge the gap, 
the number of states and the computational complexity increase exponentially as the problem scale increases. 
Recently,
owing to the powerful representation capability of deep learning, 
many deep-learning-based methods \cite{Hidasi16,Li17,Liu18} have been successfully applied to SBR. 
In particular,
some approaches \cite{Hidasi16,Li17} exploit Recurrent Neural Networks (RNNs) to characterize the item's sequential information in the session. 
However, 
these RNN-based methods are incapable of capturing long-term item dependencies. 

Recently, Graph Neural Network (GNN) \cite{Wu19,Pan20,Chen22} has attracted more and more attention due to their powerful learning ability for graph structure data representation.
To explore the complex transition relation between items in the session, GNN-based SBR constructs  sessions into graphs and utilizes GNNs to model the session graph. 
For instance, SR-GNN \cite{Wu19} first converts the session into a graph and utilizes Gated GNN \cite{Li16} to model the session graph to explore the complex transition relations between items in the session. 
After that, GC-SAN \cite{Xu19} further extends SR-GNN by adding self-attention mechanism. 
However, constructing sessions into graphs will introduce noise and lose the sequential order information, so some GNN-based methods are proposed to alleviate these problems. 
LESSER \cite{Chen20} improves the way of graph construction for sessions, taking into account the relative order of nodes in sessions. 
SGNN-HN \cite{Pan20} alleviates the long-range dependency problem by introducing a Star GNN, which improves the information propagation mechanism between items.
All these methods only focus on utilizing the internal information in a session.

\subsection{Cross-session Learning in SBR}
Utilizing the current session only to make recommendations is constrained by its limited information.
To incorporate collaborative information from external sessions, 
some collaborative filtering-based SBR methods are proposed to enhance the current session representation.
For example, 
CSRM \cite{Wang19} incorporates the relevant information contained in the neighborhood sessions by adopting a memory module to obtain more accurate session representations. 
CoSAN \cite{Luo20} utilizes multi-head attention mechanism to fuse item representations in collaborative sessions by building dynamic item representations.
GCE-GNN \cite{Wang20} simultaneously constructs local session graphs and a global graph, then extracts information related to the current session from the global graph. 
MTD \cite{Huang21} constructs a global graph connecting adjacent items in each session and utilizes graphical mutual information maximization to capture global item-wise transition information to enhance the current session's representation.
$S^2$-DHCN \cite{Xia20} utilizes hypergraph convolutional networks to capture high-order item relations and constructs two types of hypergraphs to learn information from inter- and intra-session.
The view augmentation in COTREC \cite{Xia21} enables the model to capture beyond-pairwise  relations among items across sessions.

\subsection{Multi-relation Learning in SBR}
Heterogeneous graphs have proven effective in handling information by modeling complex high-order dependencies among heterogeneous information.
They can extract user interests more accurately through global item relations across sessions \cite{Chen22,Zhang22,Han22}. 
AutoGSR \cite{Chen22} uses Network Architecture Search (NAS) techniques to automatically search for better GNN architectures that capture information on local-context relations and various item-transition semantics. 
MGIR \cite{Han22} utilizes item relations of incompatible and co-occurrence relations to generate enhanced session representations, while CoHHN \cite{Zhang22} proposes a heterogeneous hypergraph network to model price preferences. 
These works either built multiple relation graphs or used hypergraphs to model artificial features or side information as auxiliary information in modeling user actions. 
However, constructing sessions into multiple relation graphs is cumbersome.
Therefore, we propose a new approach that models association analysis of items’ categories in SBR based on a single heterogeneous graph.

\section{PRELIMINARIES}
In this section, we introduce the problem statement of SBR and the definition of item-side information. 

\subsection{Problem Statement}
We formally formulate the task of session-based recommendation (SBR).
Let \begin{math}V=\left\{v_1, v_2, \ldots, v_m\right\}\end{math} be all of items, 
where $m$ is the number of items in $V$. 
Assuming that all sessions are denoted as \begin{math}U=\left\{S_1, S_2, \ldots, S_n\right\} \end{math}, 
where $n$ is the number of sessions. 
Each anonymous session $S_\tau$ in $U$, which is denoted by 
$ S_\tau=\left\{v_1^{\tau}, v_2^{\tau}, \ldots,v_t^{\tau}\right\}$, 
consists of a sequence of interactions in chronological order, 
where $v_t^{\tau}$ denotes the item that the user interacted with at the $t$-th timestamp in the session $S_\tau$, and the length of $S_\tau$ is $t$. 
The goal of SBR is to recommend the next item from $V$ that is most probably interacted with by the user given the current session $S_\tau$. 
We call the item that interacted at the $t+1$-th timestamp the target item or the ground truth item of the session, i.e., $\left(\left[v_1, v_2, \ldots, v_{t}\right], v_{t+1}\right)$ is a session and its target item pair.

\subsection{Item-side Information}
Item-side information describes the item itself and can provide extra complementary information for the recommendation.
For each item $v_i$, we use its category $c_{v_i}$ as the item-side information to assist in learning user preference.
Let \begin{math}C=\left\{c_1, c_2, \ldots, c_l\right\}\end{math} be all of categories of items, where $l$ is the number of categories of items in $C$.
Each category of item $v_i$ is encoded into an unified embedding space, i.e., $h_{i}^c \in \mathbb{R}^d$.

\section{THE PROPOSED METHOD}
\begin{figure*}[h]
  \centering
\includegraphics[scale=0.45]{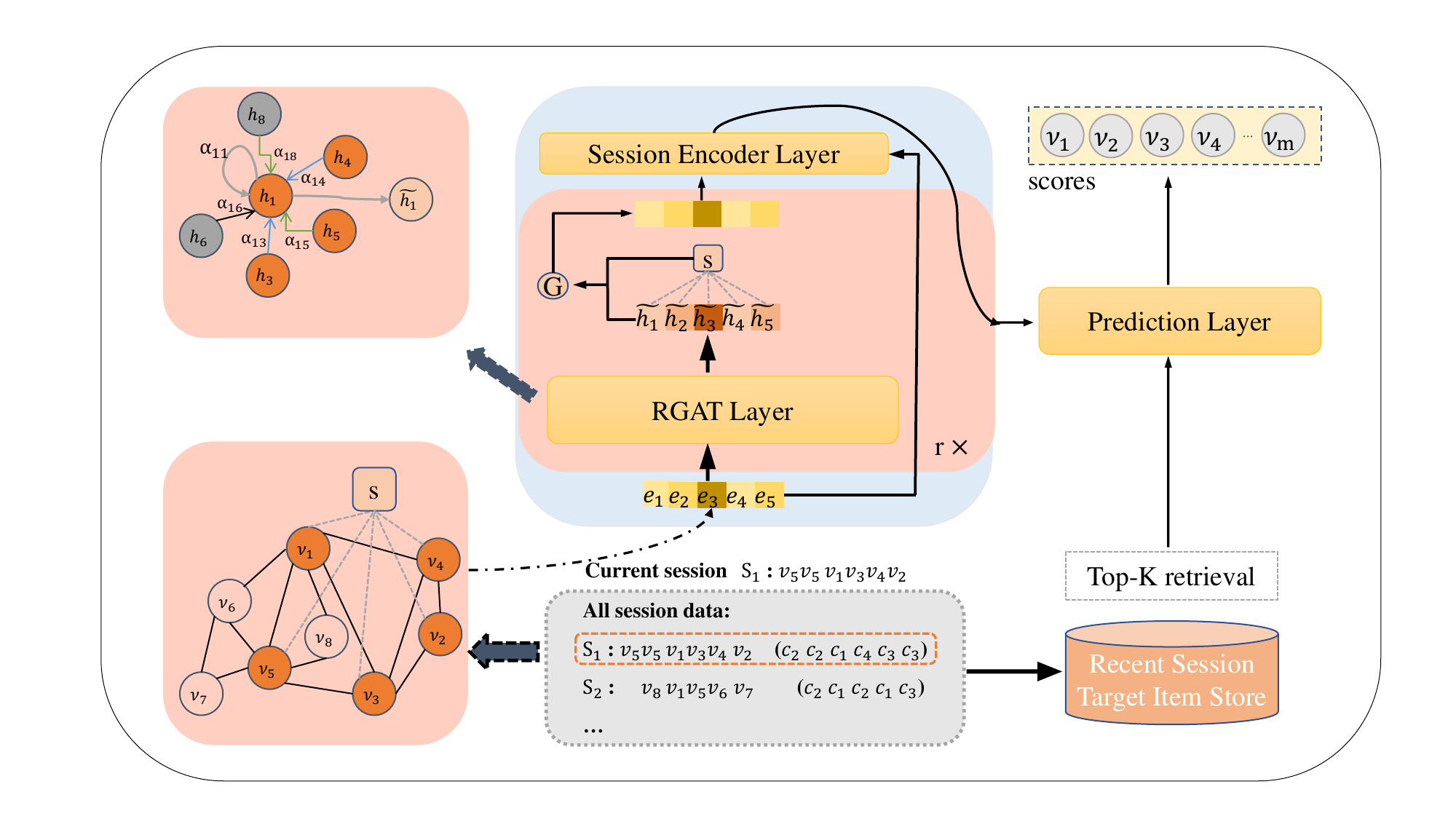}
  \caption{The overview of \name.} 
  \label{fig:model}
\end{figure*}

This section elaborates on our proposed novel Context-aware Graph Neural Networks for Session-based Recommendation (\name). 
We first give an overview of \name, which is illustrated in Figure~\ref{fig:model}.
Next, we describe each component in detail.
\subsection{Multi-relation Cross-session Graph}
Most early graph-based methods \cite{Wu19,Pan20,Xu19,Chen20,Chen22} model the item transition patterns in a single session into graphs only
and 
ignore the 
{global-level relations} between items in different sessions.
Therefore, 
we propose to build connections between different sessions to employ global-level item transition relations further.
Specifically,
we follow \cite{Wang20} to build these edge connections based on $\varepsilon$-neighbor sets of items in all sessions,
which are formally defined as follows.

\begin{definition}\label{definition:neighor}
$\varepsilon$-Neighbor set 
\cite{Wang20}. 
Given a set of sessions $U$,
for an item $v_i^x$ in session $S_x$, 
its $\varepsilon$-Neighbor set 
is a set of items with:
\begin{small} 
\begin{equation}
\mathcal{N}_\varepsilon(v_i^x)=\left\{v_j^y \middle| j\in[k-\varepsilon,k+\varepsilon], \forall v_k^y = v_i^x,
v_k^y \in S_y,
S_y \in U \right\}, \notag
\end{equation}
\end{small} 
where $i,j,k$ are positions of items in corresponding sessions, respectively. 
Further,
$\varepsilon$ is used to control 
the neighboring range of item transition.
\end{definition}

\begin{figure}[htbp]
  \centering
\includegraphics[scale=0.38]{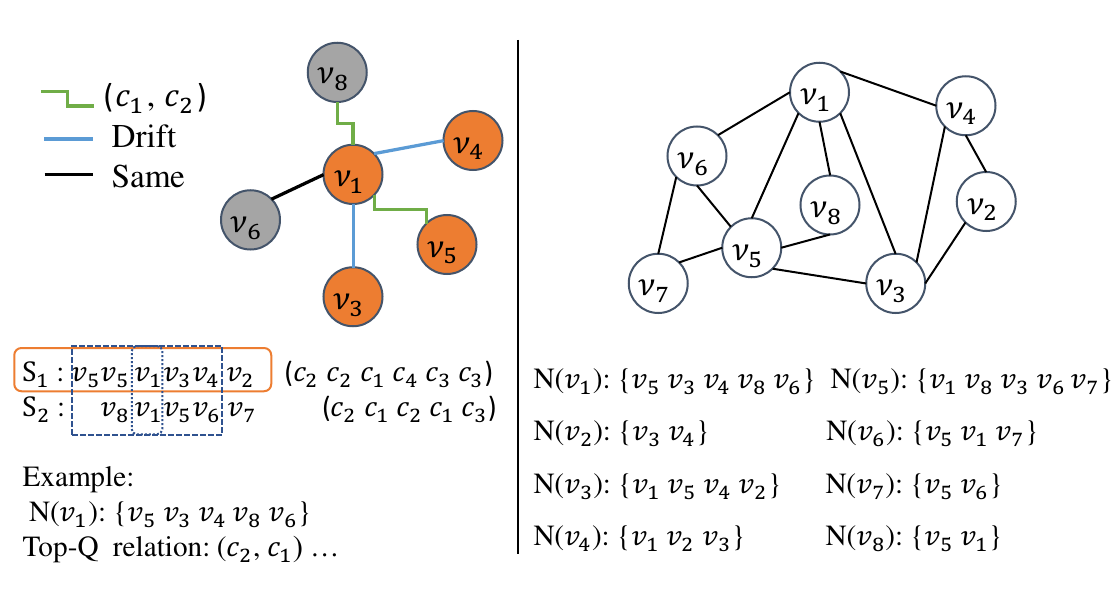}
  \caption{Illustration of the construction of the cross-session graph. Here, we set $\varepsilon$ = 2.}
  \label{fig:graph}
\end{figure}

Based on the $\varepsilon$-Neighbor set,
items from different sessions can be linked. 
Some existing works \cite{Wang20,Pan22} consider all the item transitions as one type of relation,
while we distinguish item transitions by taking the categorical attribute of items into consideration.
Intuitively, 
if a user successively clicks on items $v_1$ and $v_2$ whose categories are different, 
we cannot simply consider
these two items are related, 
because this could also indicate the drift of user interest.
Further, 
if a user sequentially clicks on items $v_1$ and $v_2$ with the same category, 
it is more likely that 
the two items 
are highly related.
This is because a user usually views a number of similar items before picking the one to buy. 
Therefore, we propose to 
construct a multi-relation cross-session graph 
based on item context and category.
Formally, 
it is defined as 
$\mathcal{G}=\left(\mathcal{V},\mathcal{E}\right)$, 
where $\mathcal{V}$ denotes the node set that contains all items in $V$ and
$\mathcal{E}=\left\{(v_i, r_{ij}, v_j) \middle| v_i\in V,v_j\in \mathcal{N}_\varepsilon(v_i) \right\}$ represents the edge set in the graph.
We use $r_{ij} = ({c_i, c_j})$ to denote the edge type, which captures the 
contextual relation
between items of categories $c_i$ and $c_j$.
Further, 
similar as in~\cite{Linden03},
for the edge 
between $v_i$ and $v_j$,
we give a weight 
{
$e_{ij}=\frac{{Freq}(v_j \in \mathcal{N}_\varepsilon(v_i))}{\left(log({Freq}(v_i)^{\alpha})+1\right)*\left(log({Freq}(v_j)^{\alpha})+1\right)}$.} 
Here,
$Freq(\cdot)$ is the frequency counting function over 
all the sessions.
To alleviate the dominant effect of a frequently occurred item,
we also introduce a hyper-parameter $\alpha$, 
whose value is set to 0.75 in our experiments.
To speed up the model efficiency,
for each item $v_i$ in the graph,
we only keep the top-$N$ neighbors in each relation that have the largest weights with it.
To further simplify the graph,
we only retain the top-$Q$ most frequent contextual relations in the graph.
For others,
we
uniformly set $r_{ij} = \texttt{Same}$ when $c_i = c_j$; $r_{ij} = \texttt{Drift}$,
when $c_i \neq c_j$.
Figure~\ref{fig:graph} shows a toy example on converting 
sessions into a multi-relation cross-session graph.

\begin{algorithm}[h]  
  \caption{The procedure of item representation learning}  
  \label{alg:Framwork}  
  \begin{algorithmic} [1]
    \Require  
    Items' initial embeddings $\textbf{h}$; The global graph $\mathcal{G}$
    \Ensure  personalized representation of the items $\textbf{h}^{s,(k)}$
    \For{each batch}
    \State sample a subgraph $\hat{\mathcal{G}}$ based on sessions in the batch
    \State $\textbf{h}^{(0)}$ index from item embedding table 
    \For{each session $[v_1,v_2,\ldots,v_t]$ in the batch}
    \State $\tilde{\textbf{h}}^{(0)}  = mean(\textbf{h}_1, \textbf{h}_2, \ldots, \textbf{h}_t)$ 
    \EndFor
    
     \For{int k=1 to L}
			\State $\textbf{h}^{(k)}=\texttt{RGAT}(\textbf{h}^{(k-1)},\hat{\mathcal{G}})$

      \For{each session in the batch}
		    \State $\delta^k=\texttt{Attention}(\textbf{h}^{(k)},\tilde{\textbf{h}}^{(k-1)})$
        \State $\textbf{h}^{s,(k)}=\texttt{Gating}(\textbf{h}^{(k)},\tilde{\textbf{h}}^{(k-1)},\delta^k)$
        \State // update node $\tilde{v}$
        \State $\beta^k=\texttt{Attention}(\textbf{h}^{s,(k)},\tilde{\textbf{h}}^{(k-1)})$
        \State $\tilde{\textbf{h}}^{(k)}=\texttt{WeightedSum}(\textbf{h}^{s,(k)},\beta^k)$
      \EndFor   
  	\EndFor
   \EndFor
  \end{algorithmic}  
\end{algorithm}

\subsection{Item Representation Learning}
\label{sec:item}
After the multi-relation cross-session graph is constructed,
we next learn item representations.
We first use an embedding look-up table to initialize embedding $\mathbf{h}_{i} \in \mathbb{R}^d$ for item $v_i$.
After that,
{we employ the attention mechanism \cite{veličković18} to generate the general embedding vector for each item based on GNNs}.
Then
we use a gating mechanism to further learn a personalized embedding vector for each item w.r.t. a given session.



\textbf{Learning General Item Representations.}
Based on the multi-relation cross-session graph,
we can easily capture both the intra-session and cross-session item-level context information.
To learn the representation of an item, 
since its $\varepsilon$-Neighbors have different importance, 
we then introduce item-level attention.
Note that for each item, it has various contextual relations. Therefore, 
when computing attention scores, 
we need to distinguish edge relations.
Specifically,
in the $k$-th layer,
the representation of item $v_i$
is derived by neighborhood aggregation,
which is formulated as:
\begin{equation}
  \begin{aligned}
{\mathbf{h}}_{i}^{(k)} = \alpha_{ii}\mathbf{W}_1\mathbf{h}_{i}^{(k-1)} +
        \sum_{v_j \in \mathcal{N}\left(v_i\right)} \alpha_{ij}\mathbf{W}_1\mathbf{h}_{j}^{(k-1)}.
  \end{aligned}
  \label{eq:agg}
\end{equation}
Here, 
the attention score $\alpha_{ij}$ is computed by
\begin{small} 
\begin{equation}
\nonumber
  \begin{aligned}
\alpha_{ij} =
        \frac{
        \exp\left(\mathbf{a}^{\top} \sigma \left( \mathbf{W}_1
        [\mathbf{h}_{i}^{(k-1)} \, \Vert \, \mathbf{h}_{j}^{(k-1)} \, \Vert \, {e}_{ij}\Vert \, \mathbf{r}_{ij}]
        \right)\right)}
        {\sum_{v_k \in \mathcal{N}\left(v_i\right) \cup \{ v_i \}}
        \exp\left(\mathbf{a}^{\top} \sigma \left( \mathbf{W}_1
        [\mathbf{h}_{i}^{(k-1)} \, \Vert \, \mathbf{h}_{k}^{(k-1)} \, \Vert \, {e}_{ik}\Vert \, \mathbf{r}_{ik}]
        \right)\right)},\
  \end{aligned}
\end{equation}
\end{small}
where $\sigma$ is the $\texttt{LeakyReLU}$
function, 
$\mathbf{a}$ and $\mathbf{W}_1$
are trainable parameters, and $\|$ denotes the {concatenation} operator. 
We also take edge weight $e_{ij}$ and edge relation embedding {$r_{ij} \in \mathbb{R}^{d}$} as edge features.



\textbf{Learning Personalized Item Representations w.r.t. Sessions.}
Note that
${\mathbf{h}}_{i}^{(k)}$ in Eq.~\ref{eq:agg} 
leverages item-level context 
and 
reflects the general embedding vector of item $v_i$.
Since an item is generally contained in various sessions,
we can further enrich the representation of an item w.r.t. a session.
Given a session $S$ and an item $v_i \in S$,
$S$ could contain many items that are not in $\mathcal{N}_\varepsilon(v_i)$
and all the items in $S$ reflect the user interest in the current session.
Therefore,
we introduce an embedding vector $\textbf{h}_i^s$
for item $v_i$
that is personalized for the session $S$.
Inspired by \cite{guo-19}, 
we add a virtual node $\tilde{{v}}$ that is 
linked to all the items in the session $S$,
whose embedding vector $\tilde{\textbf{h}}$ 
is 
used to capture the information of all the items in $S$. 
After that,
we apply a gating mechanism to fuse 
$\mathbf{h}_{i}$ and
$\tilde{\mathbf{h}}$ to
generate $\textbf{h}_{i}^{s}$:
\begin{equation}
  \mathbf{h}_{i}^{s,{(k)}}=\left(1-\delta_i\right) 
  {\mathbf{h}}_{i}^{(k)} +\delta_i \tilde{\mathbf{h}}^{(k-1)},
  \label{eq:h_new}
\end{equation}
where the gating score $\delta_i$ is computed by:
\begin{equation}
\delta_i=\operatorname{\texttt{Sigmoid}}\left(\frac{\left(\mathbf{W}_{2} {\mathbf{h}}_{i}^{(k)}\right)^{\mathrm{\top}} 
  \left(\mathbf{W}_{3} \tilde{\mathbf{h}}^{(k-1)}\right)}{\sqrt{d}}\right),
  \label{eq:his}
\end{equation}
where $\mathbf{W}_{2}$,$\mathbf{W}_{3}$ $\in \mathbb{R}^{d\times d}$ are learnable parameters,
and $\sqrt{d}$ is the scaling coefficient. 
In this way,
we can generate a personalized embedding vector $\textbf{h}_i^s$ for item $v_i$ w.r.t. the session $S$.
When $\delta_i$ is small,
$\textbf{h}_i^s$ will be close to the general representation of item $v_i$; otherwise, 
$\textbf{h}_i^s$ will be more indicative to the information in the current session $S$.
Finally,
the embedding vector $\tilde{\textbf{h}}^{(k)}$ for $\tilde{v}$ in the $k$-th layer is updated as:
\begin{equation}
  \tilde{\mathbf{h}}^{\left(k\right)} =  \sum_{v_i \in S} \mathbf{\beta}_i \mathbf{h}_{i}^{s,{(k)}},
  \label{eq:hs}
\end{equation}
where the weight ${\beta_i}$ is calculated by: 
\begin{equation}
  {\beta_i} =\operatorname{\texttt{Softmax}}\left(\frac{\left(\mathbf{W}_{4} 
  \mathbf{h}_{i}^{s,{(k)}} \right)^{\top} \left(\mathbf{W}_{5} 
  \tilde{\mathbf{h}}^{\left(k-1\right)}\right)}{\sqrt{d}}\right).
\end{equation}
Note that $\mathbf{W}_{4}$,$\mathbf{W}_{5}$ $\in \mathbb{R}^{d\times d}$ are trainable parameters. 

\subsection{Session Representation Learning}
Given a session $S$,
although the embedding of the virtual node $\tilde{{v}}$ contains the information of all the items in $S$,
it omits the temporal information and
cannot be simply taken as the representation of the session.
In the previous section,
for each item $v_i$,
we have computed its general embedding $\textbf{h}_i$ and personalized embedding $\textbf{h}_i^s$ w.r.t. a session $S$, respectively.
For notation brevity,
we overload the embedding of item $v_i$ as $\textbf{h}_i$
and next
show how to calculate session representations based on item representations.

To leverage item sequence in a session,
in addition to item embeddings,
we further incorporate 
the positional information of items and the length of the session.
For all the sessions,
we use a shared position embedding look-up table $\textbf{P}$,
where the $r$-th row $\textbf{p}_r \in \mathbb{R}^d$ represents the embedding vector for the $(t-r)$-th reverse position in a session of length $t$.
Note that we choose a reverse order for positions 
because the most recent items could be more useful for the prediction of the next item in the session.
We also introduce
a shared session length embedding look-up table $\textbf{L}$, where the $t$-th row $\textbf{l}_t\in \mathbb{R}^d$ corresponding to the embedding for the length $t$ of a session.
Note that we limit the maximum length of a session to be {$t_{max}$}.
After that,
for the $i$-th item in a session of length $t$,
we unify both the information of item position and session length into the item embedding $\textbf{h}_i$, 
and output an updated embedding $\mathbf{z}_i$ 
for $v_i$:
\begin{equation}
  \mathbf{z}_i= \mathbf{h}_{i} + \mathbf{p}_{t-i} + \mathbf{l}_t.
  \label{eq:z_i}
\end{equation}


To calculate the representation of a session,
we can also employ item categories in the session.
For all the items,
we further define a shared item category embedding look-up table, where
each row indicates an embedding vector of an item category.
Given a session $S$ of length $t$,
we unify item categories in the session as:
\begin{equation}
  \mathbf{h}_c=\texttt{Mean}(\{ \mathbf{h}_i^c \}_{i=1}^t, \forall v_i \in \mathcal{S}),
\end{equation}
where $\mathbf{h}_i^c$ represents the category embedding of the item $v_i$.
Then
we use the attention mechanism to fuse the information of all the items in $S$,
and have:
\begin{equation}
  \bar{\mathbf{z}}_s=\sum_{i=1}^t \gamma_i \mathbf{z}_i,
\end{equation}
where
the attention weight $\gamma_i$ can be calculated by a two-layer MLP:
\begin{equation}
\gamma_i= \texttt{MLP}(
        \mathbf{z}_{i} \, \Vert \, \mathbf{z}_{t}\, \Vert \, \tilde{\mathbf{h}} \, \Vert \, \mathbf{h}_c)
\end{equation}
Here, 
$\tilde{\mathbf{h}}$ is the embedding of the virtual node $\tilde{v}$ in Equation~\ref{eq:hs}.
we also use the embedding $\textbf{z}_t$ of the last item in $S$ because it could be highly related to the prediction of the next item.

After that, 
we combine 
$\bar{\mathbf{z}}_s$
and $\mathbf{z}_t$ to capture user interests in session $S$:
\begin{equation}
  \mathbf{z}_s=\mathbf{W}_6\left[\bar{\mathbf{z}}_s \Vert \mathbf{z}_t\right],
  \label{eq:zs}
\end{equation}
where 
$\mathbf{W}_6$ is a weight parameter.
Further,
inspired by the skip connection technique in \cite{Xu18}, 
we directly derive embedding of item $v_i$ from the look-up table and rerun Equations~\ref{eq:z_i}-\ref{eq:zs} to generate a new $\textbf{z}_s$ (we denote it as $\textbf{z}_s'$ for difference) without the item representation learning stage in Section~\ref{sec:item}.
Finally, 
the representation of session $S$ is computed by:
\begin{equation} 
 \mathbf{h}_s=\mathbf{z}_s + {\mathbf{z}_s'}.
\end{equation}


\subsection{Label Collaboration}
Most existing works \cite{Wu19,Wang19,Wang20,Luo20,Pan20} use the one-hot encoded vector of the target item as the hard label of user preference, which may not reflect the true preference.
The intuition is that users are generally only exposed to a limited number of items, 
so the lack of other items 
could induce a bias to user interest.
Further,
user preference is also influenced by different time periods and contextual scenarios,
which can deviate from historical data over time.
Therefore, 
to address the problem,
we employ the session-level contexts and
propose a label collaboration strategy,
which aims to explicitly utilize the target items of historical sessions with most similar behavioral patterns to the current session as collaborative label information. 


\textbf{Collaborative Sessions Retrieval.}
Given a session $S$, 
our target is to first retrieve 
$K$ sessions that are most similar to $S$ from a fixed-size candidate session pool with {$M$} most recent sessions.
Intuitively, the more sessions we retrieve,
the more accurate the user preference could be estimated,
and the larger computation cost will be induced.
Therefore,
we further utilize  \emph{SimHash}~\cite{Charikar02} 
to 
speed up the model efficiency. 
The SimHash function takes the session representation as input and generates its binary fingerprint,
where
each entry 
is either 0 or 1.
It has been pointed out in \cite{Chen21} that 
the outputs of SimHash satisfy the \emph{locality-sensitive properties} that the outputs are similar if the input vectors are similar to each other.
Specifically, 
we first project embeddings of $S$ and other candidate sessions
into binary fingerprints by {multiplying} the input embedding vectors with a hash function, 
which is set to be a fixed random projection matrix {$H \in \mathbb{R}^{d\times m}$},
where $m<d$.
As a result, similar session embedding vectors can get the same hashing output.
After that,
we calculate the \emph{hamming distance} between the output vectors 
and select 
the top-$K$ most similar sessions to $S$ 
from $M$ candidate sessions by: 
\begin{equation}
\nonumber
  N_{S},W_{S} = \texttt{topK}\left(-\texttt{HammingDistance}\left(\mathbf{e},\hat{\mathbf{e}}\right)\right), 
\end{equation}
where 
$\mathbf{e} = \texttt{SimHash}({S})$, $\hat{\mathbf{e}} = \texttt{SimHash}(\hat{{S}})$, 
and $\hat{S}$ is derived from $M$ candidate sessions.
The weights $W_{S}$ are then normalized to ensure that they sum to 1.
We denote the set of one-hot encoded labels of selected sessions as $N_{S}=\left\{\textbf{y}_1^{S}, \textbf{y}_2^{S}, \ldots, \textbf{y}_K^{S}\right\}$
and the set of corresponding weights as
$W_{S}=\left\{w_1^{S},w_2^{S}, \ldots, w_K^{S}\right\}$,
which will be used for label 
collaboration of session $S$.
Further,
the pool is updated by a slide window scheme:
removing the oldest sessions and 
adding the most recent ones in the next batch.
Therefore,
compared to the time complexity $\mathcal{O} \left(MBd\right)$ of retrieval by cosine similarity in~\cite{Wang19}, the time complexity of our retrieval is {$\mathcal{O} \left(Bm\right)$}, where $B$ is the batch size, 
$M$ is the pool size and 
$m$ is smaller than session representation dimensionality $d$.

\textbf{Collaborative Label Generation.}
After $K$ most similar sessions are retrieved,
we next construct the soft label for session $S$.
These $K$ sessions can help provide more comprehensive estimation for user interests than using $S$ only.
Therefore,
we 
obtain the collaborative label for $S$ by a weighted sum of the one-hot encoded label of each retrieved session:
\begin{equation}
\tilde{\mathbf{y}}=\sum_{i=1}^K   w_i^{S}\textbf{y}_i^{S}.
\end{equation}

\subsection{Prediction Layer}
The prediction layer is used to output the probability distribution of items that the user will interact at the next timestamp in the current session. 
Due to the long-tail distribution problem \cite{Gupta19} in the data for recommendation, 
we normalize 
item embeddings and session  embeddings in each layer.
Finally, we feed them into a prediction layer, where 
the inner product and the \texttt{Softmax} function are applied to generate the output:
\begin{equation}
  \hat {\mathbf{y}}_i=\texttt{Softmax}({\mathbf{h}}_s^\top {\mathbf{h}}_{i}),
\end{equation}
where $\hat{\mathbf{y}}_i$ denotes the probability of interacting with item $v_i$ 
in the next timestamp.
The total loss function consists of two components: 
a cross-entropy loss based on the hard label $\mathbf{y}$ and a KL-divergence loss based on the soft label $\tilde{\mathbf{y}}$:
\begin{equation}
  \mathcal{L} =\textrm{CrossEntropy}(\hat{\mathbf{y}}, \mathbf{y})+\lambda \textrm{KLD}(\hat{\mathbf{y}}, \tilde{\mathbf{y}}),
\end{equation}
where 
$\lambda $ is a trade-off parameter that is used to control the
importance of the two components.

\section{EXPERIMENTS}
In this section, 
we conduct extensive experiments on three 
publicly available datasets
to show the effectiveness of our method.
We preprocess these datasets as in 
\cite{Wu19}.
First, 
we arrange all the sessions 
in the chronological order and split the data into training data and test data by the timestamps of sessions. 
Second, we filter out items that appear less than 5 times or only appear in the test set, and also
the sessions of length one. 
Third, 
we perform data augmentation with a temporal-window shifting to generate more data samples in a session,
e.g., 
$\left(\left[v_1, v_2, \ldots, v_{n-1}\right], v_n\right),\ldots,\left(\left[v_1, v_2, \right], v_3\right),\left(\left[v_1 \right], v_2\right)$ for session $\left[v_1, v_2, \ldots, v_n\right]$.
Further,
we adopt two widely used evaluation metrics in information retrieval: \emph{Precision (P@20)} and 
\emph{Mean Reciprocal Rank (MRR@20)} for evaluating the 
performance. 

\begin{table}[htbp]
\centering
  \caption{datasets statstics}
  \label{tab:stats}
  \begin{tabular}{c|c|c|c}
    \hline \hline Dataset & Diginetica & Tmall & Yoochoose1\_64 \\
    \hline  \#Train sessions & $719,470$ & $351,268$ & $369,859$ \\
    \#Test sessions & $60,858$ & $25,898$ & $55,898$ \\
    \#Items & $43,097$ & $40,728$ & $16,766$ \\
    Avg. lengths & $5.12$ & $6.69$ & $6.16$ \\
    \hline \hline
    \end{tabular}
 \end{table}
 
\subsection{Datasets} 
The following datasets are utilized to evaluate our model.
The statistics of the processed datasets are shown in Table~\ref{tab:stats}.

$\bullet$ \textbf{Diginetica}\footnote{http://cikm2016.cs.iupui.edu/cikm-cup} contains anonymous user transaction information extracted from e-commerce search engine logs for five months. The dataset is from CIKM Cup 2016.

$\bullet$ \textbf{Tmall}\footnote{https://tianchi.aliyun.com/dataset/dataDetail?dataId=42} records the anonymized users' shopping logs on the online shopping platform called Tmall. The dataset comes from the IJCAI15 competition.

$\bullet$ \textbf{Yoochoose1\_64}\footnote{http://2015.recsyschallenge.com/challege} was built by YOOCHOOSE GmbH to support RecSys Challenge 2015. It records users' clicks from an e-commerce website. 
We follow Wu \cite{Wu19} by using the most recent proportion $1/64$ of the training sessions.

\subsection{Hyper-parameter Setup}
Following \cite{Wu19,Pan20}, the dimension of the latent vectors is fixed to 256, and the batch size is set to 100.
We use the Adam optimizer with the initial learning rate of 0.001, which will decay by 0.8 after every 3 epochs. 
The $l_2$ penalty is set to $10^{-5}$ and the dimension of the hash matrix in SimHash is set to 64.
The candidate number of sessions is set to 1500 in the label collaboration strategy.
We set the parameter $\lambda$ for adjusting the loss weights to 0.1 for Deginetica 5 for Yoochoose1\_64, and 10 for Tmall.
We vary the number of retrieved target items in label collaboration from $\left\{10, 30, 50, 70, 90\right\}$ and the number of frequent contextual relations from $\left\{0,5,10,20,30\right\}$ to study their effects.

\subsection{Baselines}
\begin{table*}[h]
\centering
  \caption{Overall performance comparison on three datasets.
  For fairness, we directly report the results of baseline methods from their original papers, where ``-'' indicates the absence of corresponding results in the original papers.}
  \label{tab:overall}
  \setlength{\tabcolsep}{4mm}{
  \begin{tabular}{ccccccc}
      \hline \hline & \multicolumn{2}{c}{ Diginetica } & \multicolumn{2}{c}{ Tmall } & \multicolumn{2}{c}{ Yoochoose1\_64 } \\
      \cline { 2 - 7 } Method & P@20 & MRR@20 & P@20 & MRR@20 & P@20 & MRR@20 \\
      \hline POP & $1.18$ & $0.28$ & $2.00$ & $0.90$ & $6.71$ & $0.58$ \\
      Item-KNN & $35.75$ & $11.57$ & $9.15$ & $3.31$ & $51.60$ & $21.81$ \\
      FPMC & $22.14$ & $6.66$ & $16.06$ & $7.32$ & $45.62$ & $15.01$ \\
       GRU4Rec & $30.79$ & $8.22$ & $10.93$ & $5.89$ & $60.64$ & $22.89$ \\
      NARM & $48.32$ & $16.00$ & $23.30$ & $10.70$ & $68.32$ & $28.63$ \\
      STAMP & $46.62$ & $15.13$ & $26.47$ & $13.36$ & $68.74$ & $29.67$ \\
      SR-GNN & $50.73$ & $17.59$ & $27.57$ & $13.72$ & $70.57$ & $30.94$ \\
      LESSR & $51.71$ & $18.15$ & $23.53$ & $9.56$ & $70.05$ & $30.59$ \\
      SGNN-HN & \underline{$55.67$} & \underline{$19.45$} & $-$ & $-$ & \underline{$72.06$} & \underline{$32.61$} \\
      
      \hline
       CSRM & $48.49$ & $17.13$ & $29.46$ & $13.96$ & $-$ & $-$ \\
      CoSAN & $51.97$ & $17.92$ & $32.68$ & $14.09$ & $-$ & $-$ \\
      GCE-GNN & $54.22$ & $19.04$ & $33.42$ & $15.42$ & $70.91$ & $30.63$ \\
      $S^2$-DHCN & $53.18$ & $18.44$ & $31.42$ & $15.05$ & $-$ & $-$ \\
      MTD  &    $51.82$ & $17.26$ & $29.12$ & $13.73$ & $71.88$ & $31.32$ \\
      COTREC &   $54.18$ & $ 19.07$ & $ 36.35$ & \underline{$ 18.04$} & $-$ & $-$ \\
      
      \hline AutoGSR & $54.56$ & $19.20$ & $33.71$ & $15.87$ & $71.77$ & $31.02$ \\
      MGIR  &    $-$ & $-$ & \underline{$36.41$} & $17.42$ & $-$ & $-$ \\
      
      \hline 
      \name\_ns & $55.29$ & $21.04$ & $38.17$ & $17.79$ & $71.82$ & $33.05$ \\
      \name & $\mathbf{56.49 }$ & $\mathbf{23.22}$ & 
      $\mathbf{38.77 }$ & $\mathbf{18.37 }$ & 
      $\mathbf{72.21 }$ & $\mathbf{34.40 }$ \\
      Improv. & $1.47\%$ & $19.30\%$ & $6.48\%$ & $1.82\%$ & $0.20 \%$ & $5.48\%$ \\
      \hline \hline
  \end{tabular}}
  \end{table*}
  
To verify the performance of our proposed model, we compared our model with 17 other methods, which can be grouped into three categories. 
Readers are referred to Section~\ref{sec:related} for more details.

\textbf{(Single Session methods):} 
\textbf{POP} recommends the most popular items.
\textbf{Item-KNN} \cite{Sarwar01} recommends items based on the cosine similarity between items in the current session and candidate items.
\textbf{FPMC} \cite{Rendle10} uses both Markov chain and Matrix Factorization to consider the user's personalized and general information.
\textbf{GRU4REC} \cite{Hidasi16} exploits the memory of GRUs by characterizing the entire sequence.
\textbf{NARM} \cite{Li17} and \textbf{STAMP} \cite{Liu18} further utilize attention mechanism additionally, which aims to capture the current interest and general interest of the user.
\textbf{SRGNN} \cite{Wu19}, \textbf{LESSER} \cite{Chen20}, \textbf{SGNN-HN} \cite{Pan20}, convert each session into a graph and do not utilize cross-session information.

\textbf{(Cross Session methods):} 
\textbf{CSRM} \cite{Wang19} incorporates the relevant information in the neighborhood sessions through the memory network.
\textbf{CoSAN} \cite{Luo20} utilizes multi-head attention mechanism to build dynamic item representations by fusing item representations in collaborative sessions.
\textbf{GCE-GNN} \cite{Wang20} and \textbf{MTD} \cite{Huang21} simultaneously focus on cross-session and intra-session dependencies.  
\textbf{COTREC} \cite{Xia21} and \textbf{$S^2$-DHCN} \cite{Xia20} employ a global argumentation view of items to mine informative self-supervision signals.

\textbf{(Multi-relation methods):}
\textbf{AutoGSR} \cite{Chen22} and \textbf{MGIR} \cite{Han22} both learn multi-faceted item relations to enhance session representation.
Note that MGIR utilizes cross-session information while AutoGSR does not.
 
\subsection{Overall performance}
From the experimental results on the three datasets in Table~\ref{tab:overall}, we have the following observations:
(1) It is observed that methods utilizing RNNs or attention mechanisms perform better than early methods such as Item-KNN and FPMC because they are both suitable for dealing with sequential data with temporal information without losing the internal-session-level context.
Methods such as CSRM and CoSAN offer higher performance for introducing auxiliary information from historical sessions than single session methods like GRU4Rec, NARM and STAMP.
This confirms the effectiveness of leveraging external-session-level contexts.
The current best-performing methods such as SGNN-HN, COTREC and MGIR are GNN-based approaches because GNNs are good at capturing complex item-transitions across sessions, 
which shows the effectiveness of introducing cross-session item-level context by graph modeling.

(2) \nameinitem outperforms other GNN-based models SR-GNN, LESSER, AotoGSR, and SGNN-HN. 
This is because all these methods are designed for local sessions without considering cross-session information in the global view. 
While the cross-session method COTREC leverages self-supervision for enhancing session representation, it ignores heterogeneity and is outperformed by \nameinitem.

(3) The leading performance of \nameinitem and COTREC over 
GCE-GNN implies that it is useful to capture the internal-session-level context in the global graph because the latter only considers the cross-session item-level context of item-transitions and lacks diversity in its collaborative information. 
Therefore, 
COTREC employs self-supervised learning to impose a divergence constraint on global view and internal-session view of item embedding, 
while \nameinitem further introduces personalized item representation w.r.t sessions.
This demonstrates the significance of the internal-session-level context in global graph modeling.

(4) Our approach achieves the best performance in all the datasets, which shows the importance of making full use of contexts in sessions.
Further, our model has a significant improvement in terms of MRR@20 on Diginetica and Yoochoose1\_64, 
indicating that the item relevant to users' interests can be ranked higher, which is critical for user experience improvement and confirms the superiority of our model.

(5) To ensure a fair comparison, we conducted experiments with an additional variant model that does not use side information to construct the graph.
As shown in Table~\ref{tab:overall}, even without utilizing the side information of the item’s category (aka \name\_ns), our method still performs well across different datasets.

\subsection{Ablation Study}
\begin{figure*}[htbp]
\centering  
\includegraphics[scale=0.38]{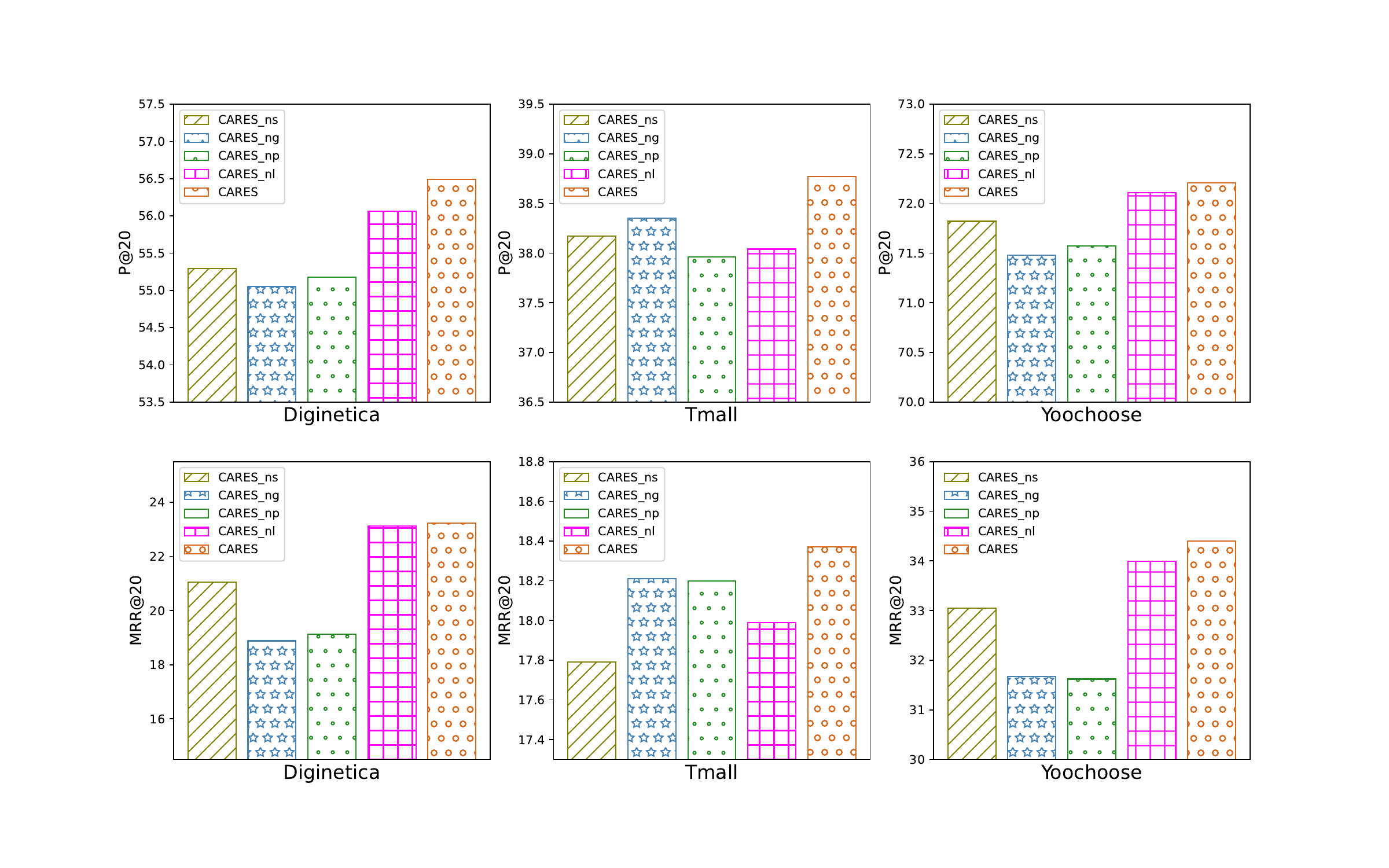}
\caption{Model performance in the ablation study}   
\label{fig:comp}   
\end{figure*}
We conduct an ablation study on \nameinitem to understand the characteristics of its main components. 
One variant updates items’ embeddings by directly capturing information from intra-session without utilizing general information to model item-transition relationships on the global graph. 
This helps us understand the importance of including cross-session item-level context in SBR.
We call this variant \textbf{\name\_ng} (\textbf{n}o \textbf{g}eneral information). 
Another variant learns items' embedding without personalized information w.r.t sessions. 
We call this variant \textbf{\name\_np} (\textbf{n}o \textbf{p}ersonalized information), which helps us evaluate the effectiveness of internal-session-level context. 
To show the importance of the label collaboration strategy, we train the model with cross-entropy loss only and call this variant \textbf{\name\_nl} (\textbf{n}o \textbf{l}abel collaboration).
\textbf{\name\_ns} (\textbf{n}o \textbf{s}ide information) represents the variant of \nameinitem without considering category information of items to understand the effect of items' category association in SBR.

From the experimental results in Figure~\ref{fig:comp}, the following observations are made.
(i) Compared with \textbf{\name\_ng}, \nameinitem leverages cross-session item-level context and thus can utilize diverse collaborative information from the global graph and outperform \textbf{\name\_ng}.
(ii) It can also be observed that \nameinitem with learning personalized information beats \textbf{\name\_np} on all the datasets. 
This indicates that internal-session-level context can effectively preserve user intent through adding personalized information w.r.t sessions.
(iii) \nameinitem performs better than \textbf{\name\_nl}, and this indicates that utilizing the target items of historical sessions with similar behavioral patterns to the current session as external-session-level context can mitigate the bias in the user preference distribution.
(iv) \nameinitem also defeats \textbf{\name\_ns}, indicating that items’ category plays an important role in learning users’ preferences.
Additionally, although side information improves recommendation accuracy, our model still performs well without it, as shown in Table~\ref{tab:overall}.

\begin{figure}[htbp]
  \centering
\includegraphics[scale=0.35]{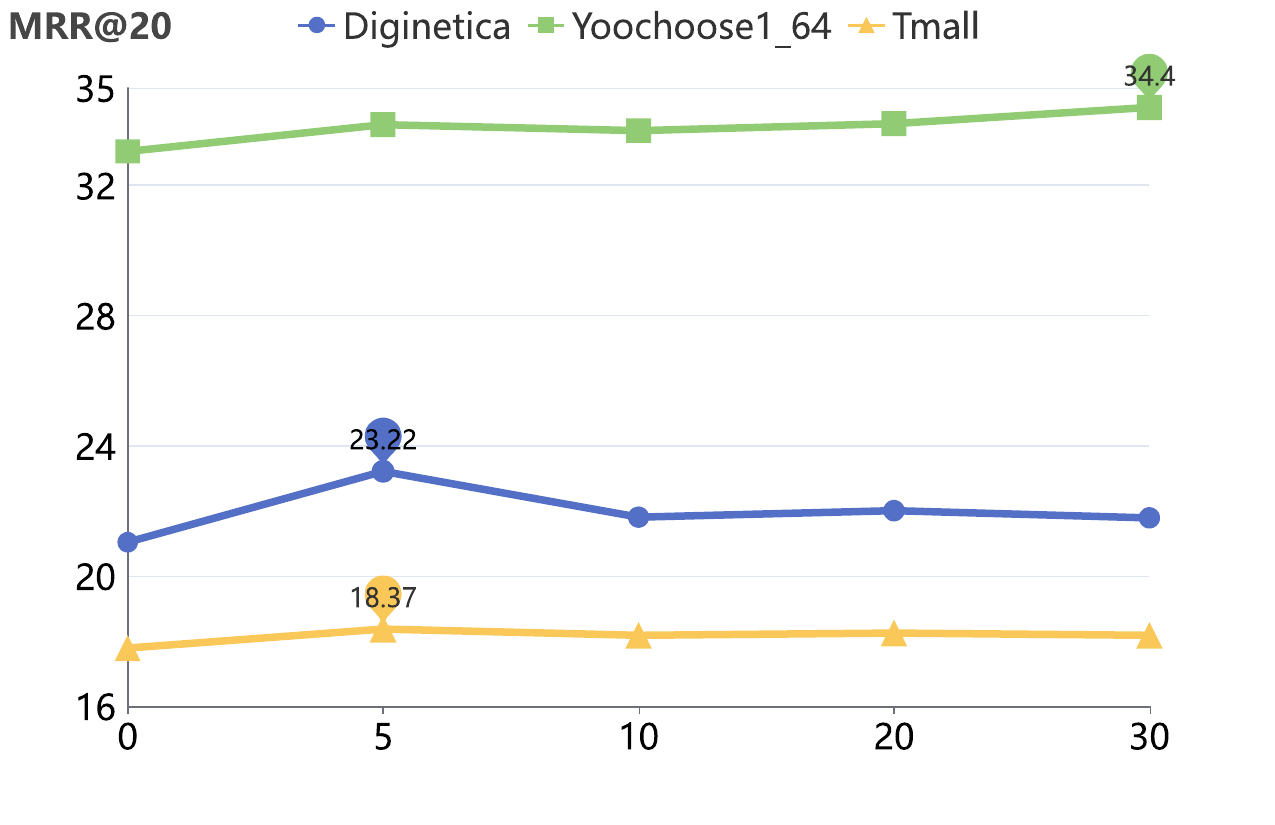}
  \caption{Performance comparison on the number of contextual relations}
  \label{fig:rels}
\end{figure}

\subsection{Influence of Contextual Relations}
In this section, we study how contextual relations affect the performance of the proposed method. 
Due to the limited space, we only show the results in terms of MRR@20. 
The results are shown in Figure~\ref{fig:rels}.
From the results, we can see that the models that do not use contextual relations always have lower performance.
This is because contextual relations can help the model capture more complex item context, which indicates disentangling the relation semantics of sessions is a promising direction for further exploiting the information across sessions.
For different datasets, the optimal number of contextual relations is different.
For the dataset Yoochoose1\_64, the score hits the highest when the relation number is set to 30.
For the other two datasets, the optimal relation number is 5 and we can see that increasing the number of relations does not always result in a better performance.
This is because only the relation between items' categories with enough high frequency can be considered a context.

\subsection{Sensitivity Analysis of Hyper-Parameters}
\begin{figure}[htbp]
  \captionsetup[sub]{skip=0pt}
  \captionsetup{skip=1pt}
  \centering
  \subcaptionbox{m}{\includegraphics[width=0.48\linewidth]{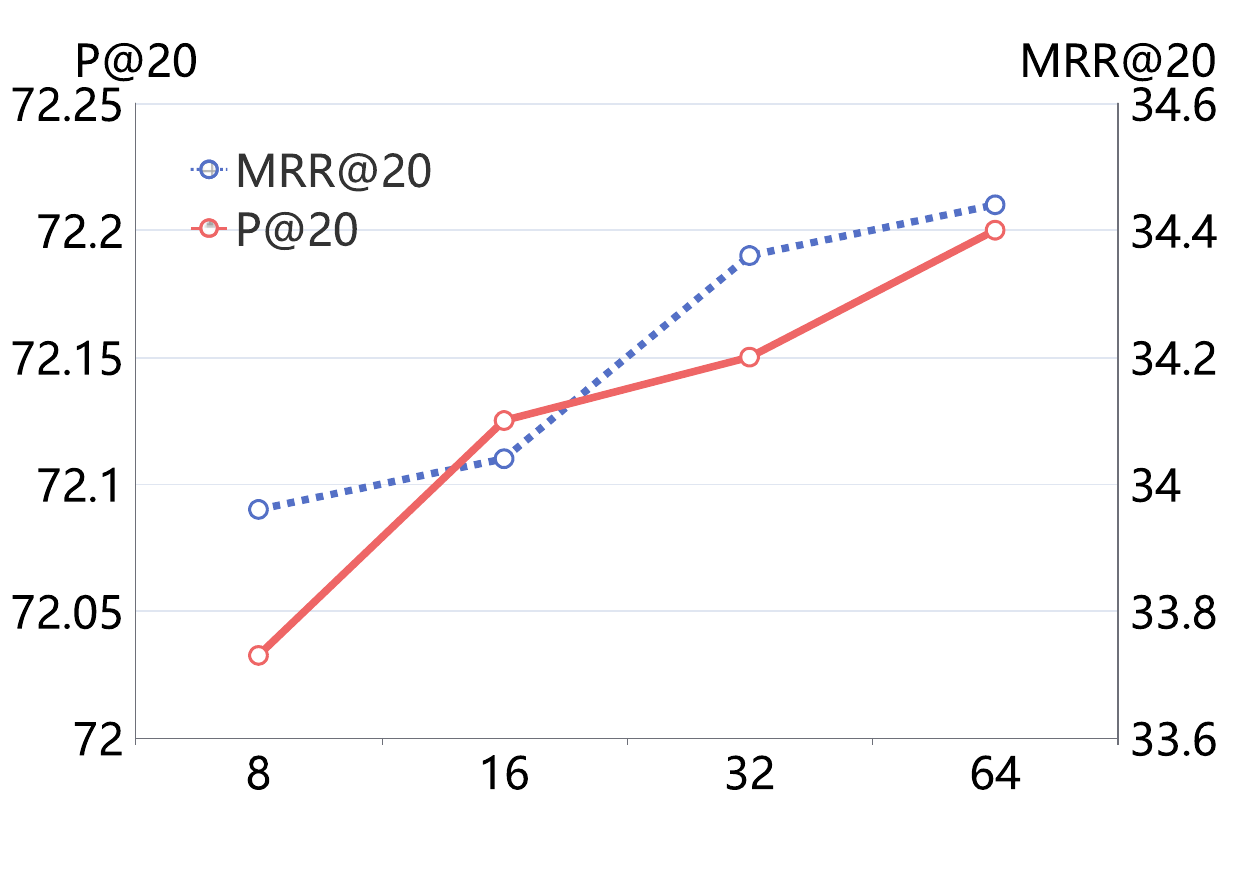}}
  \subcaptionbox{K}{\includegraphics[width=0.48\linewidth]{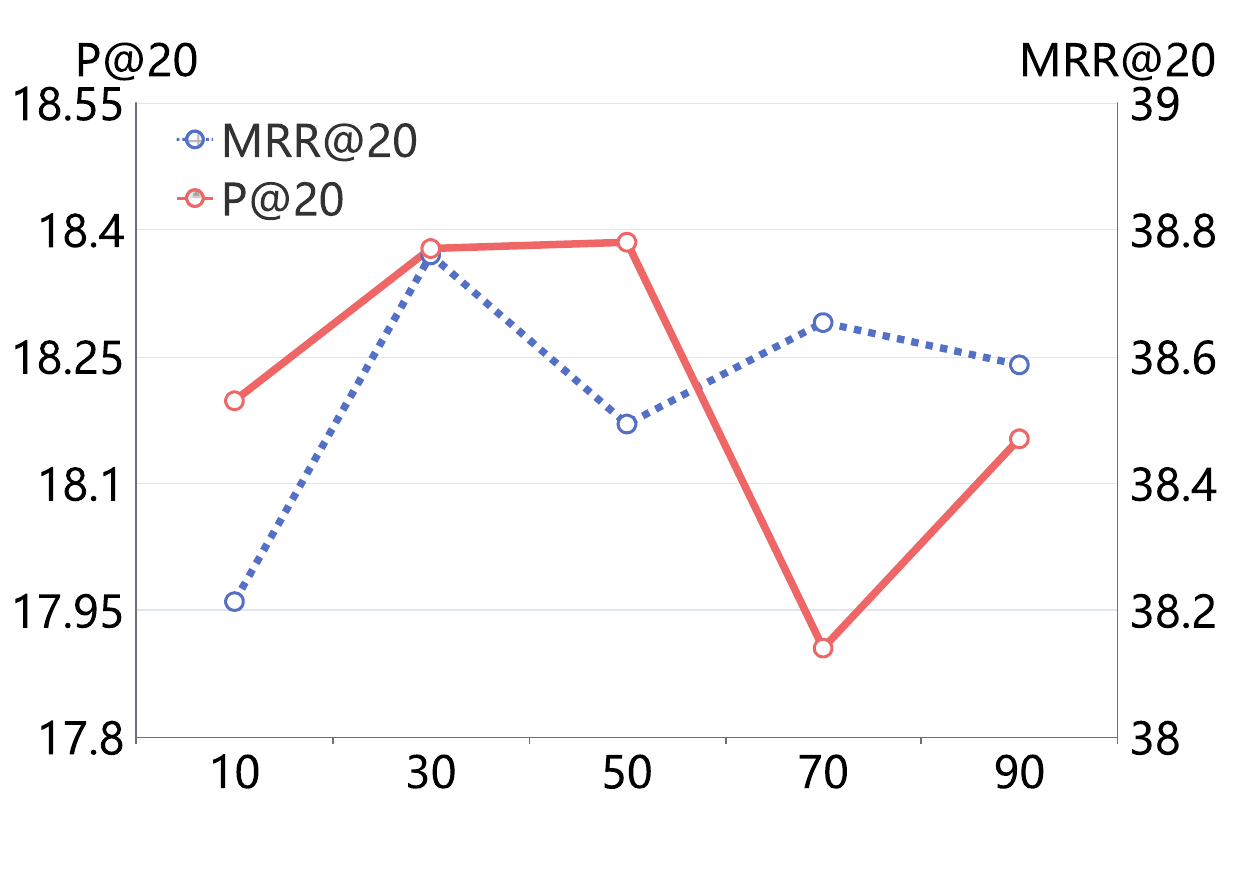}}
  \caption{Sensitivity Analysis of Hyper-Parameters}
  \label{fig:hyper}
\end{figure}

We end this section with a sensitivity analysis on the hyper-parameters of \name. 
In particular, we study two hyper-parameters: the hash matrix dimension $m$ and the number of retrieved sessions $K$.
In our experiments, we vary one parameter each time with others fixed. 
Fig~\ref{fig:hyper} illustrates the results with w.r.t. P@20 and MRR@20 scores on the datasets of Tmall and Yoochoose1\_64. 
(Results on other datasets scores exhibit similar trends, and thus are omitted for space reasons.) 
From the figure, we see that

(1) A larger dimension $m$ can slightly improve the performance of the model. 
Since the model is not very sensitive to the hash matrix dimension, setting a small size of $m$ can also guarantee the performance of the model.

(2) Fewer retrieved sessions in label collaboration are not sufficient to provide enough information for the current session. 
And there is also a performance drop when retrieving more sessions, which shows that a large number of collaborative sessions could contain noise that adversely affects the recommendation performance.
So, an appropriate number of retrieved sessions $K$ is essential.

\section{CONCLUSION}
In this paper, we propose a novel method named \nameinitem for session-based recommendation based on graph neural network.
Specifically, it converts the session sequences into a global graph with item attributes as context.
The general item representations are generated by various contextual relations through item-level attention.
After that, we apply a gating mechanism to further enrich the representations of items with personalized information w.r.t sessions.
Then the intra- and cross-session context information are subsequently combined to enhance the recommendation performance.
Finally, it incorporates label collaboration to generate soft user preference distribution as labels and thus empowers the proposed model to alleviate the overfitting problem.
Comprehensive experiments demonstrate that our proposed model can make full use of contexts in sessions, especially those cross-session ones, thus achieving state-of-the-art performance over three real-world datasets consistently.

\vspace{12pt}

\end{document}